\documentclass[twocolumn,
aps,nofootinbib,showpacs,showkeys,preprint
tightenlines,preprintnumbers,] {revtex4}

\usepackage{epsf,epsfig,subfigure,graphicx,amsmath,amssymb}
\usepackage{amsfonts}
\usepackage{latexsym}
\usepackage{color}

\newcommand{\OVER}[1]{\,\overline{\hskip -0.5mm #1}}

\newcommand{\dis}[1]{\begin{equation}\begin{split}#1\end{split}\end{equation}}

\newcommand{\etal}{et al.}

\newcommand{\tev}{\,\textrm{TeV}}
\newcommand{\gev}{\,\textrm{GeV}}

\newcommand{\Z}{{\mathbb Z}}

\newcommand{\Xz}{{X^{(0)}}}
\newcommand{\Xzc}{{\,\overline{\hskip -0.5mm X}^{(0)\, }}}

\newcommand{\MGUT}{M_{\rm G}}

\newcommand{\fiveb}{\overline{\bf 5}\,}
\newcommand{\five}{{\bf 5}}
\newcommand{\tenb}{\overline{\bf 10}\,}
\newcommand{\ten}{{\bf 10}}
\newcommand{\one}{{\bf 1}}

\newcommand{\ie}{{\it i.e.~}}

\newcommand{\Higgs}{{\it Higgs-flavor-democracy}}

\begin{document}
\draft

\title{\Large\bf Natural Higgs-flavor-democracy solution of the $\mu$  problem of
    supersymmetry and the QCD axion
}

\author{Jihn E.  Kim}
\affiliation
{Department of Physics, Kyung Hee University, Seoul 130-701, Korea
}

\begin{abstract}
We show that the hierarchically small $\mu$ term in supersymmetric theories is a consequence of two identical pairs of Higgs doublets taking a democratic form for their mass matrix. We briefly discuss the discrete symmetry $S_2\times S_2$ toward the democratic mass matrix. Then, we show that there results an approximate Peccei-Quinn symmetry and hence the value $\mu$ is related to the axion decay constant.

\end{abstract}


\maketitle

\section{Introduction}

The large hadron collider (LHC) at CERN in Geneva, Switzerland was designed to study the mass scale problem in physics, and has produced enough data over the last year to conclude an important discovery \cite{Higgs126}. Also, the recent reports from satellite experiments \cite{Planck1, AMS13} have attracted a great deal of attention regarding possible particle spectrum in the TeV (or 100--1000 GeV) mass region. Probably, the satellite data will select in the near future some theoretical models in the TeV energy region. Search for all the particles in the Standard Model\,(SM) has been completed with the Higgs boson discovery \cite{Higgs126}. The Higgs boson is responsible for giving masses to all the SM particles and the discovery has indeed confirmed the mechanism giving mass, with a remarkable precision, to gauge bosons \cite{HiggsMech} and to fermions \cite{Weinberg67}. In the SM, this is possible by introducing just one Higgs doublet: $\{H^+,H^0\}$ or  $\{H^0,H^-\}$.

Fundamental particles influence the evolution of the universe. Among these, light particles, light in the Planck mass unit $M_P $\,($\simeq 2.44\times 10^{18}\,\gev$), dominated the recent evolution dynamics of the universe. The new Alpha Magnetic Spectrometer\,(AMS) data does not rule out the possibility that a cold dark matter  component is weakly interacting massive particle with mass near TeV \cite{AMS13}. The recent reports \cite{Planck1} from the Planck collaboration support the inflationary idea on the density perturbation without a noticeable non-gaussianity. The inflationary idea is based on general relativity with an inflaton field(s) which is a `fundamental-scalar' boson. Indeed, the discovery of Higgs boson as a fundamental scalar supports the `fundamental-scalar' inflaton idea and influences already in selecting possible allowed inflationary models \cite{PeirisBonn}. In this Letter, we are interested in two fundamental scalars, the Higgs boson and the very light axion \cite{KSVZDFSZ}.

But the light `fundamental-scalar' from the Planck mass or from the grand unification\,(GUT)
mass\,($\MGUT\simeq 2.5\times 10^{16}\,\gev$) point of view is the so-called gauge hierarchy problem. Because of the quadratic divergence of the `fundamental-scalar' mass, an extreme fine-tuning of boson mass parameters is needed and now, after the discovery of the Higgs boson, we have the real `fundamental-scalar' mass problem. Supersymmetry\,(SUSY) is an attractive idea mildly extending the SM to reliably control the `fundamental-scalar' mass problem.
The simplest extension is called the Minimal Supersymmetric Standard Model (MSSM), introducing a superpartner for every SM particle except for the Higgs doublet.

\begin{figure}[!t]
\begin{center}
\includegraphics[width=1\linewidth]{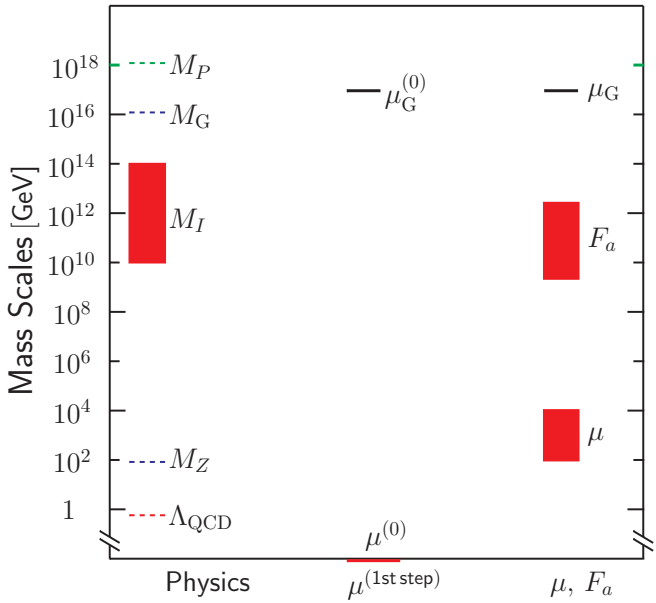}
\end{center}
\caption{Important mass scales in particle physics. The scale $\mu^{(0)}$ is for the lowest order toward a natural solution, and the scale $\mu$ is generated at higher orders. Here, $M_I$ the intermediate scale. } \label{fig:MassScales}
\end{figure}

For the Higgs doublet, it is extended to one pair $H_u \equiv \{H_u^+,H_u^0\}$ and $H_d \equiv \{H_d^0,H_d^-\}$ and then supersymmetrized. In the MSSM, however, there are two serious problems: the $\mu$ problem \cite{KimNilles84} and the strong CP problem \cite{PQ77}. The $\mu$ term, $\mu H_uH_d$, gives order $\mu$ Higgs boson mass. The problem is that this $\mu$ term breaks no low energy symmetry and is naturally expected to be of order the Planck scale $M_P$ or the  GUT scale $\MGUT$.\footnote{We note that the $\mu$ problem has been discussed long time ago with the PQ symmetry \cite{KimNilles84}, with supergravity effects \cite{Giudice88} and in string models \cite{CasasMunozMu93}.  More recently, the discrete $Z_4^R$ symmetry has been discussed also \cite{LeeRatz11}.
}
In Fig. \ref{fig:MassScales} in the first column, we show several interesting hierarchically separated mass scales of particle physics.

The strong CP problem is a problem on how the QCD vacuum angle $\theta$ is bounded to the extremely small region, $|\theta|<10^{-10}$ \cite{InvAxionRev10}. The most attractive solution is extending the SM again with a symmetry, \ie via the Peccei-Quinn (PQ) global symmetry \cite{PQ77}, eventually leading to the invisible axion \cite{KSVZDFSZ} and axino by SUSY \cite{KimSeoAxino12}. However, the PQ global symmetry is not warranted in  gravitational interactions \cite{GravityQCDaxion}. Even though no superpartner of the SM has not been found so far, still the SUSY extension seems most attractive for the `light\,(\ie 126\,GeV)' Higgs boson, possibly with SUSY manifested somewhat above TeV, for example by the family dependent U(1) quantum numbers \cite{KimJE12}. So we continue to use SUSY here.

To understand small masses, we anticipate a scenario shown in the second column of Fig. \ref{fig:MassScales} where zero mass is obtained in the first step. The Goldstone boson idea may work for this. However, due to the gravity dilemma on global symmetries \cite{GravityQCDaxion}, we do not adopt a global symmetry. Surprisingly, we find that if `democracy' among the same type of particles is present, some massless particles result, and even some approximate global symmetries can follow. In order to address the democracy problem with SUSY, we must spell out the $\mu$ problem. On this road, we obtain an approximate global symmetry which will be interpreted as an approximate PQ symmetry. Thus, the gravity argument against axion \cite{GravityQCDaxion} does not apply here since we never introduce the PQ symmetry. So finally, the axion decay constant is related to the SUSY parameter $\mu$, and sets a scale for the axion decay constant $F_a$ at the intermediate scale $M_I$.

Obtaining massless particles naturally was considered long time ago for three families of fermions right after the discovery of $\tau$ and $b$ under the name of `flavor democracy' \cite{demoform}. The chief motivation of Ref. \cite{demoform} was in order to obtain the heavy third family fermions, rather than to obtain two massless fermions of the first two families. For the flavor democracy of three up-type quarks, one introduces the permutation symmetries on three flavors, independently for the left-handed(L) fields and for the right-handed(R) fields: $S_3(L)\times S_3(R)$. Then, the mass matrix for the up-type quarks takes the form,
\dis{
\left(\begin{array}{ccc}{m_t}/{3}~ &{m_t}/{3}~  &{m_t}/{3}\\[0.5em]
 {m_t}/{3}~ &{m_t}/{3}~  &{m_t}/{3}\\[0.5em]
{m_t}/{3}~ &{m_t}/{3}~  &{m_t}/{3} \end{array}\right)\,,  \nonumber
}
leading to the mass eigenvalues of $m_t,\, 0,$ and 0. This kind of $S_3\times S_3$ is not helpful for obtaining just one pair of massless Higgs doublets, or by SUSY one pair of massless Higgsino doublets.

Therefore, to obtain just one massless pair of Higgsino doublets in SUSY models, we need $S_2(\tilde{H}_u)\times S_2(\tilde{H}_d)$ for two Higgsino doublet pairs, $\{\tilde{H}_u^{(i)}\,, \tilde{H}_d^{(i)}\}$ with $i=1,2$. Here, $S_2(\tilde{H}_u)$ is the interchange symmetry $(1\leftrightarrow 2)$ for two Higgsino doublets $\tilde{H}_u^{(1)}$ and $\tilde{H}_u^{(2)}$. So is $S_2(\tilde{H}_d)$ for two Higgsino doublets $\tilde{H}_d^{(1)}$ and $\tilde{H}_d^{(2)}$. Due to  SUSY, we identify the Higgsino discrete symmetry $S_2(\tilde{H}_u)\times S_2(\tilde{H}_d)$ with the Higgs boson discrete symmetry $S_2({H}_u)\times S_2({H}_d)$, shortened as \Higgs. If we start with just one pair of Higgs doublets, we cannot understand the {\it fundamental problem of $\mu$} naturally.

\section{Higgs-flavor-democracy}

Our solution of the $\mu$ problem is by introducing just two pairs of Higgsino doublets at the GUT scale $\MGUT$. But, the unification of gauge coupling constants at a GUT scale $\MGUT$ requires only one pair of Higgs doublets $\{H_u,\, H_d\}$ at low energy \cite{GUTunif81}, which is understood as in the second column of Fig. \ref{fig:MassScales} with just one pair surviving below the GUT scale. If the two Higgsino pairs are not distinguished by any quantum number and geometry of the internal space, there must be the permutation symmetries, $S_2(H_u)$ and $S_2(H_d)$.
Then, the democratic form for the Higgsino mass matrix can be obtained,
\dis{
\left(\begin{array}{cc} \MGUT/2\,,\, &  \MGUT/2\\  \MGUT/2\,,\,  & \MGUT/2  \end{array}\right)\label{eq:demo2}
}
which gives Higgsino mass eigenvalues  $ \MGUT$ and 0.
The \Higgs~ based on $S_2(H_u)\times S_2(H_d)$ works as follows. The representation content of $S_2$ is only a singlet  $\one$. The tensor product (for the mass matrix) of two $S_2$ singlets, \ie $H_u=({H}_u^{(1)},{H}_u^{(2)})^T$ and $H_d=({H}_d^{(1)},{H}_d^{(2)})^T$, contains two independent real parameters $m_E$ and $m_O$ for the diagonal and off-diagonal combinations, respectively,
\dis{
W  &=  \Big[m_E ({H}_u^{(1)\alpha}{H}_d^{(1)\beta}+{H}_u^{(2)\alpha}{H}_d^{(2)\beta})
\\
&+ m_O ({H}_u^{(1)\alpha}{H}_d^{(2)\beta}+{H}_u^{(2)\alpha}{H}_d^{(1)\beta})\Big]
 \epsilon_{\alpha\beta}\label{eq:WS2}
}
where $\alpha$ and $\beta$ are the SU(2)$_W$ gauge group indices of the SM. Now, let us apply $S_2(H_u)$ symmetry to $W$: $1 \to 2, 2\to 1$ for $H_u^{(i)}$, and $1 \to 1, 2\to 2$ for $H_d^{(i)}$. Then, we obtain
\dis{
W \to  &\Big[ m_E ({H}_u^{(2)\alpha}{H}_d^{(1)\beta}+{H}_u^{(1)\alpha}{H}_d^{(2)\beta})\\
&+ m_O ({H}_u^{(2)\alpha}{H}_d^{(2)\beta}+{H}_u^{(1)\alpha}{H}_d^{(1)\beta})\Big]
 \epsilon_{\alpha\beta}\,.\label{eq:WS2perm}
}
Comparing Eqs. (\ref{eq:WS2}) and (\ref{eq:WS2perm}), we obtain $m_E=m_O$ and obtain the \Higgs, Eq. (\ref{eq:demo2}). Applying $S_2(H_d)$ gives the same result. But if we apply $S_2(H_u)$ and $S_2(H_d)$ simultaneously, $m_E$ and $m_O$ are not related.

An $S_2(H_u)\times S_2(H_d)$ invariant superpotential can be tried with
\dis{
W_{S_2\times S_2}= \frac{\MGUT}{2} &\Big({H}_u^{(1)\alpha}{H}_d^{(1)\beta}+{H}_u^{(2)\alpha}{H}_d^{(2)\beta}
 + {H}_u^{(1)\alpha}{H}_d^{(2)\beta}+{H}_u^{(2)\alpha}{H}_d^{(1)\beta}\Big)
 \epsilon_{\alpha\beta}\, \label{eq:WS2S2}
 + \frac{M_\epsilon}{2} \left({H}_u^{(1)\alpha}{H}_d^{(1)\beta}+ {H}_u^{(2)\alpha}{H}_d^{(2)\beta}\right).
}
Note, however, that the $M_\epsilon$ term alone has an additional continuous symmetry SO(2).
Restricting to the discrete symmetry only, we set $M_\epsilon=0$.
The mass matrix is diagonalized to the new $(H^{(0)}, H^{(\rm G)})^T$ basis,
\dis{
& M^{\rm Higgsino}_0=\left(\begin{array}{cc} 0 & 0\\ 0&~~\MGUT
\end{array}\right),\\
& H_{u,d}^{(0),(\MGUT)}=\frac{1}{\sqrt2}\left(H_{u,d}^{(1)}\mp H_{u,d}^{(2)}\right).\label{eq:diagbases}
}

Indeed, one can find a few string models allowing two identical pairs of Higgs doublets  in the MSSM \cite{Ovrut05TwoH,KimKyaeMSSM}. We note that Ref. \cite{KimKyaeMSSM} contains two pairs of Higgs doublets in the twisted sector $T_6$, has the \Higgs, and so naturally contains a light pair of Higgs doublets.

\section{Generation of TeV scale $\mu$}

Since the \Higgs~ gives one pair of the Higgsino doublets zero mass, one has to break the \Higgs~ to obtain a TeV scale $\mu$, or the massless Higgsino can never obtain mass.
In the SUSY field theory framework, we show a possibility that the \Higgs~ is broken. Let us take the minimal K\"ahler potential $K=\Phi_i\Phi_i^\dagger$
where $\Phi_i\,(i=1,2)$ is the gauge group non-singlet field such as the Higgs superfield and $X_i\,(i=1,2)$ and $\,\overline{\hskip -0.5mm X}_i\,(i=1,2)$ are gauge group singlet superfields, obeying the common $S_2\times S_2$ symmetry of $\Phi_i\,(i=1,2)$ and  $\overline{\Phi}_i\,(i=1,2)$,
\dis{ S_2:
~\Phi_1 \leftrightarrow \Phi_2 ,~ X_1\leftrightarrow X_2\,,\label{eq:CommonS2}
}
and similarly for the barred fields.
Let us introduce a very light QCD axion in the SM singlet fields \cite{KSVZDFSZ}, $\Xz$ and $\Xzc$, for $ 10^{9}\,\gev\simeq F_a\simeq 10^{12}\,\gev$. For the VEVs of  $\Xz$ and $\Xzc$ not to be fine-tuned, the moduli from antisymmetric tensor $B_{MN}$ are not suitable for the QCD axion since the corresponding decay constants are above $10^{16}\,\gev$ \cite{ChoiKim85}. So, we assume that $X$ and $\,\overline{\hskip -0.5mm  X}$  are arising from some matter representations which are nontrivial representations of a subgroup of $E_8$, for example. $H_{u,d}$, $X$ and $\,\OVER{X}$ are Higgs matter and they can be put in the same representations of a GUT group, e.g. $(\times,\times,\times, H_u^+, H_u^0, X,\cdots)^T$ and  $(\times,\times,\times, H_d^0, H_d^-, \,\OVER{X},\cdots)^T$. In this example, $S_2(L)$ acts commonly on $H_u$ and $X$, and $S_2(R)$ acts commonly on $H_d$ and $\,\OVER{X}$. Then, $XX$ and $\,\overline{\hskip -0.5mm X} \,\overline{\hskip -0.5mm  X}$ are not allowed and we consider only $X \,\overline{\hskip -0.5mm  X}$.
Let us consider the following $S_2(L)\times S_2(R)$ symmetric non-renormalizable term \cite{KimNilles84},
\dis{
W^{\rm (nonren.)}&=\sum_{i,j=1,2}\left(\frac{X^{(i)}\,\overline{\hskip -0.5mm  X}^{\,(j)}}{M_P}\right)H_u^{(i)}H_d^{(j)}\\
&+\sum_{ij}  \sum_{kl} \left(\frac{X^{(i)}\,\overline{\hskip -0.5mm  X}^{\,(j)}}{M_P'}\right)H_u^{(k)}H_d^{(l)}\,.
\label{eq:KNterm}
}
We have not included both $X_1\OVER{X}_1$ and $X_2\OVER{X}_2$ terms because in general $X_i$ in $(\times,\times,\times, H_u^+, H_u^0, X_i)^T$ and $\OVER{X}_i$ in $(\times,\times,\times, H_d^0, H_d^-, \,\OVER{X}_i)^T$, being not moduli, can have different extra U(1) quantum numbers. Also, there exists a string argument disallowing them \cite{Casas93}.
Both terms of (\ref{eq:KNterm}) combine to parametrize interactions of the light fields. For the effective interaction of light fields, the fields with index $^{(0)}$ matter, and detailed combinations are hidden and we will present it with the first term only.
To have VEVs of $X$ and $\,\overline{\hskip -0.5mm X}$ fields, let us consider an $S_2\times S_2$ symmetric superpotential with a singlet $Z\,(Z\to Z~{\rm under}~ S_2)$ \cite{Kim84},
\dis{
W\propto Z\left(X_1\,\overline{\hskip -0.5mm X}_1+X_1\,\overline{\hskip -0.5mm X}_2+X_2\,\overline{\hskip -0.5mm X}_1+X_2 \,\overline{\hskip -0.5mm X}_2 - F_a^2\right)\,.\label{eq:S2FieldTh}
}
Here, we removed the tadpole term of the $X_i$ and $\,\overline{\hskip -0.5mm X}_i$ fields by an appropriate matter parity such as $P(X_i)=P(\,\overline{\hskip -0.5mm X}_i)=-1$, and others with the even parity.
There exists a flavor-democracy breaking minimum, $\langle Z\rangle =0 , ~\langle X_1\rangle = ~\langle \,\overline{\hskip -0.5mm X}_1\rangle = F_a , ~\langle X_2\rangle = \langle \,\overline{\hskip -0.5mm X}_2\rangle =0$. Since there also exists the $S_2\times S_2$ symmetric vacuum $\langle Z\rangle =0 , ~X_1=\,\overline{\hskip -0.5mm X}_1=X_2=\,\overline{\hskip -0.5mm X}_2\ne 0$, our choice of democracy breaking minimum is spontaneous.  At the democracy breaking vacuum, $\langle X_1\rangle=\langle \,\overline{\hskip -0.5mm X}_1\rangle =F_a$ and $\langle X_2\rangle = \langle X_2\rangle=0$, we generate the following term
\dis{
W^{\rm (nonren.)}= \frac{\lambda\,X_1  \,\overline{\hskip -0.5mm X}_1}{2M_P} (H_u^{(0)}+H_u^{(\MGUT)}) (H_d^{(0)}+H_d^{(\MGUT)})
   \,.
\label{eq:KNnumer}
}
From Eqs. (\ref{eq:diagbases}) and (\ref{eq:KNnumer}), we obtain the following Higgsino mass matrix
\dis{
M^{\rm (Higgsino)} =\left(\begin{array}{cc} \mu &  \mu \\ \mu &~~\MGUT+ \mu
\end{array}\right)
}
where $\mu=F_a^2/2M_P$. The eigenvalues of $M^{\rm (Higgsino)}$ are $\mu-\frac{\mu^2}{\MGUT}$ and $\MGUT+ \mu(1+ \frac{\mu}{\MGUT})$. Choosing $F_a$ at the intermediate scale $\sim 10^{10-12}\,\gev$, we obtain the TeV scale $\mu$.

\section{Invisible axion}

\begin{figure}[!t]
\begin{center}
\includegraphics[width=0.95\linewidth]{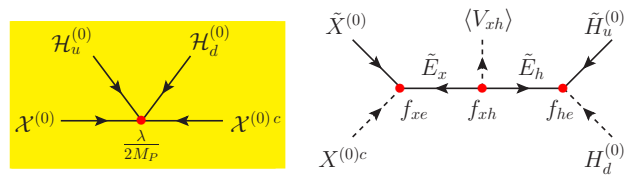}
\end{center}
\caption{The shaded diagram for the effective dimension-4 superpotential of superfields ${\cal X}^{(0)}$ and ${\cal H}^{(0)}$. This generates the $\mu$ term and defines the PQ charges. It can arise from a GUT completed model shown on the RHS where $\tilde{E}_{x,h}$ are the GUT scale inos.
} \label{fig:dim5}
\end{figure}

\begin{figure}[!t]
\begin{center}
\includegraphics[width=0.5\linewidth]{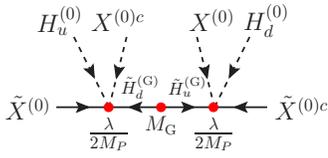}
\end{center}
\caption{A diagram violating the PQ symmetry. Here, $M_G$ is a GUT scale Higgsino mass.
} \label{fig:PQviolation}
\end{figure}

In the $S_2\times S_2$ breaking vacuum, Eq. (\ref{eq:KNnumer}), integrating out the heavy fields
$H_{u,d}^{(\MGUT)}$, we have the light field interaction
\dis{
W = \frac{\lambda \Xz \Xzc }{2M_P} H_u^{(0)} H_d^{(0)}
\label{eq:KNlightPQdef}
}
which was anticipated below Eq. (\ref{eq:KNterm}).
The Higgs multiplets $ H_u^{(0)}$ and $H_d^{(0)}$ couple to quarks, $W= -q u^c H_{u}^{(0)} -q d^c H_{d}^{(0)}$, and define their PQ charges. Then, the PQ charges of $X^{(0)}$ and $\,\overline{\hskip -0.5mm X}^{\,(0)}$ are given through Eq. (\ref{eq:KNlightPQdef}). The diagram Fig. \ref{fig:PQviolation} breaks the PQ symmetry, gives a correction to $\mu$, which  is smaller than Fig. \ref{fig:dim5} by a factor $\mu/\MGUT$. Thus, the PQ symmetry is approximate, and the explicit PQ symmetry breaking term considering $H_{u,d}^{(\MGUT)}$ will lead to a very small $\theta_{\rm QCD}$ term at the order $\mu/\MGUT\sim 10^{-14}$, well below the current bound of $10^{-10}$ \cite{InvAxionRev10}.
But the SUSY breaking must be considered, otherwise the axion is massless due to the massless gluino.
Then, the soft term(the A-term from Fig. \ref{fig:PQviolation}) breaks the PQ symmetry with the strength,
\dis{
m&_{3/2} \frac{\lambda^2}{4 M_P^2} \left(\frac{1}{M_G}H_u H_d \right) (XX^c)^2=\frac{\lambda^2 m_{3/2}v_u v_d F_a^4}{8M_P^2 M_G}\\
&\approx \left(\frac{\lambda^2 }{\tan\beta}\right)3\times 10^{-4}\, \left(\frac{m_{3/2}}{\tev}\right)\left(\frac{\mu}{\tev}\right)^2\,[\gev^4]
}
where $\tan\beta=v_u/v_d$.
To have $|\bar\theta|<10^{-9}$, we need the extra contribution to the axion potential is no more than $4\times 10^{-12 }\,\gev^4$.  So, in the gravity mediation scenario we need $ \lambda^2 /\tan\beta\lesssim O(10^{-8})$, or $ |\lambda|\lesssim O( 10^{-4}\sqrt{\tan\beta})$. One may guess that $\lambda$ of Fig. \ref{fig:dim5} is 1 if it appears from the gravity sector, but it is not so. The $1/M$ coupling for example arises from the RHS diagram of Fig. \ref{fig:dim5}, so that
$f_{xe}f_{he}/f_{xh}V_{xh}$ leads to $\lambda=f_{xe}f_{he}M_P/M_{\tilde E}$ where $M_{\tilde E}=f_{xh}V_{xh}$. Taking $M_P/M_{\tilde{E}}\approx 10$, we need a product of Yukawa couplings of $|f_{xe}f_{he}|\lesssim 10^{-5}\sqrt{\tan\beta}$. So, $f_{xe}$ and $f_{he}$ each of order $10^{-2}$ will satisfy this bound. In the gauge mediation scenario, the gravitino
mass is in the eV range and the resulting $|\,\overline{\hskip -0.5mm  \theta}|$ is completely negligible.

Note that the VEV of $X_1$ (and $\overline{X}_1$) is the common scale for breaking the $S_2$ symmetry toward a nonzero $\mu$ and the axion decay constant $F_a$, anticipated long time ago in \cite{Kim84}, to realize the invisible axion \cite{KSVZDFSZ}. Probably, it is a good rationale that the invisible axion scale is the intermediate scale $M_I$. So, the axion mass is in the range $10\,\mu{\rm eV}-1\,{\rm meV}$  \cite{InvAxionRev10}. It has been observed that gravity is not respecting the PQ global symmetry \cite{GravityQCDaxion}, and hence the components of the antisymmetric tensor field $B_{MN}$ in string models \cite{MDaxion} and approximate global symmetries from string \cite{AxionMatter}  were tried for the QCD axion. So, the QCD axion based on the discrete symmetry even at the field theory level is circumventing all these worries. The effective PQ symmetry we obtain here from the matter field $X_1$ and $\overline{X}_1$ leads to a reasonable QCD axion from string.

\section{On the GUTs}

Finally, we comment on how the solution of the $\mu$ problem is resolved in GUTs.
For a complete GUT example, we refer to the flipped SU(5) GUTs \cite{flipSU5} from the $\Z_{12-I}$ orbifold compactification \cite{KimKyaeZ12,HuhKK09} and from the fermionic string \cite{fermionFlip}. In particular, we find two pairs of Higgs doublets in the twisted sector $T_4^0$ of Ref. \cite{HuhKK09}, as shown in Table \ref{table:T40}. But we present the discussion at the  supersymmetric field theory level below. The GUT Higgs multiplets $h_u^{(i)}\equiv \five_{-2}^{(i)}$ and $h_d^{(i)}\equiv \fiveb_{2}^{(i)}$ have the infamous doublet-triplet splitting problem. In the flipped SU(5), it is resolved by coupling the color (anti-)triplets in $h_u$ and $h_d$ to the color (anti-)triplets in $\,\overline{\hskip -0.5mm H}\equiv\tenb_{-1}$ from $T_3$ and $H\equiv\ten_{1}$ from $T_9$ by $W\sim\sum_{i} HHh_u^{(i)}+\sum_i\,\overline{\hskip -0.5mm H} \,\overline{\hskip -0.5mm H} h_d^{(i)}$ \cite{fermionFlip}.\footnote{For the fields of Table \ref{table:T40}, one should attach singlets from the twisted sector $T_2^0$ of Ref. \cite{HuhKK09} to obtain the factors in $W$.}
But one combination out of two color triplet pairs of $h_u$ and $h_d$ remains massless if $S_2(h_u)\times S_2(h_d)$ is unbroken with the same reason for the case of \Higgs. Including the color (anti-)triplets in $\,\overline{\hskip -0.5mm H}$ and $H$ and color (anti-)triplets in two sets of $h_u^{(i)}$ and $h_d^{(i)}$, we expect the following mass matrix for the color triplets,
\begin{table}[!t]
\begin{center}
\begin{tabular}{|c|c|c||c|}
\hline Sector & Weight &
Mult. & $SU(5)_X$ \\
\hline\hline
$T_{4}^0$ &$\left(\underline{1\,0\,0\,0\,0};\frac{1}{3}\, \frac{1}{3}\, \frac{1}{3}
\right)(0^8)'$ &   $2$  & $ \five_{-2}\, (h_u^{(i)})$ \\
$T_{4}^0$ &$\left(\underline{-1\,0\,0\,0\,0};\frac{1}{3}\, \frac{1}{3}\, \frac{1}{3}
\right)(0^8)'$ &   $2$   & $ \fiveb_{2}\, (h_d^{(i)})$ \\[0.3em]
\hline
$T_{4}^0$ &$\left({0^5};\frac{-2}{3}\, \frac{-2}{3}\, \frac{-2}{3}
\right)(0^8)'$ &    $3$  & $ \one_0\,(X_i)$ \\[0.3em]
\hline
$T_{3}$ &$\left(\underline{\frac12\,\frac12\,\frac12\,\frac{-1}2\,\frac{-1}2}; 0^3 \right)\left(0^5;\frac{-1}{4}\,\frac{-1}{4}\,\frac{2}{4}\right)'$ &    $1$  & $\tenb_{-1}\,
(\overline{H})$ \\
$T_{9}$ &$\left(\underline{\frac12\,\frac12\,\frac{-1}2\,\frac{-1}2\,\frac{-1}2}; 0^3 \right)\left(0^5;\frac{1}{4}\,\frac{1}{4}\,\frac{-2}{4}\right)'$ &    $1$  & $\ten_{1}\,(H)$
 \\[0.4em]
\hline
$T_{2}^0$ &$\left(0^5\,; \frac{-1}3\, \frac{-1}3\, \frac{-1}3 \right)\left(0^5;\frac{-1}{2}\,\frac{1}{2}\,0\right)'$ &    $1+1$  & $\one_0$ \\
$T_{2}^0$ &$\left(0^5\,; \frac{-1}3\, \frac{-1}3\, \frac{-1}3  \right)\left(0^5;\frac{1}{2}\,\frac{-1}{2}\,0\right)'$ &    $1+1$  & $\one_0$
 \\[0.4em]
\hline
\end{tabular}
\end{center}
\caption{The $T_4^0, T_3, T_9$, and $T_2^0$ left-handed states discussed in the text. The flipped SU(5) quantum numbers are those of SU(5)$\times$U(1)$_X$.}\label{table:T40}
\end{table}

\dis{
M_{\rm color}\propto \left(\begin{array}{ccc} \delta &\xi_1 & \xi_2\\
\eta_1 & \MGUT & \MGUT \\
\eta_2 & \MGUT & \MGUT
\end{array}\right)\nonumber
}
where $\delta,\xi_i$ and $\eta_i$ are in general of order $\MGUT$.
If $S_2(h_u)\times S_2(h_d)$ remains unbroken, we have $\xi_1=\xi_2$ and $\eta_1=\eta_2$ and Det.$\,M_{\rm color}=0$, implying one massless pair of color triplet and anti-triplet. For the $S_2(h_u)$ symmetry, we can break it by $W_{h_u}=\lambda_1  HHh_u^{(1)}+\lambda_2  HHh_u^{(2)}$ with $\lambda_1\ne \lambda_2$, and similarly for the $S_2(h_d)$ symmetry by $W_{h_d}$. With $\lambda_1$ and $\lambda_2$ of order 1, all color trplets and anti-triplets become superheavy.
In the flipped SU(5), one cannot break the \Higgs~ for the Higgs doublets via $W_{h_u}$ and $W_{h_d}$ because $\langle H\rangle$ and $\langle \overline{H}\rangle$ do not give mass to the Higgs doublets \cite{fermionFlip}. Thus, introducing $W_{h_u}$ and $W_{h_d}$, we achieve the doublet-triplet splitting in the flipped SU(5) GUT.

\section{Conclusion}

Introducing a global symmetry in string models toward the strong CP solution by spontaneous breaking of the PQ symmetry has been a dilemma for a long time. In this Letter, we have found a mechanism to introduce it approximately on the way to solve the $\mu$ problem with \Higgs. The underlying symmetry is the discrete $S_2(H_u)\times S_2(H_d)$ symmetry for two identical pairs of Higgs doublets, $\{H_u^{(i)},H_d^{(i)}\}\,(i=1,2)$, at the high energy scale. Being discrete, the $S_2(H_u)\times S_2(H_d)$ symmetry can be realized in string models.
In sum, it has not escaped our attention that two identical pairs of Higgs doublets with supersymmetry introduce \Higgs, bring one pair down to the TeV scale solving the $\mu$ problem naturally, and predicts a very light axion. Finally, we note that the underlying discrete symmetry is free from the gravity argument against the axion.

\section*{Acknowledgments}
I thank Bumseok Kyae for useful discussions. This work is supported in part by the National Research Foundation (NRF) grant funded by the Korean Government (MEST) (No. 2005-0093841).




\begin{thebibliography}{99}
\bibitem{Casas93} Note that the $Z_3$ orbifold compactification allows only cubic and higher order terms \cite{CasasMunozMu93}.
Our prototype example, the fields of $T_4^0$ twisted sector of $Z_{12-I}$ behave equivalently to those of $Z_3$.
\end{thebibliography}
\end{document}